\documentclass{ws-mpla}
\usepackage[super]{cite}
\usepackage{graphicx}
\begin{document}

\markboth{H.W. Ang et. al.}
{Modifed combinant analysis of the $e^+e^-$ multiplicity distributions}

%%%%%%%%%%%%%%%%%%%%% Publisher's Area please ignore %%%%%%%%%%%%%%
\catchline{}{}{}{}{}
%%%%%%%%%%%%%%%%%%%%%%%%%%%%%%%%%%%%%%%%%%%%%%%%%%%%%%%%%%%%%%%%%%%

\title{Modifed combinant analysis of the $e^+e^-$ multiplicity distributions}

\author{\footnotesize H. W. Ang\footnote{corresponding author}, M. Ghaffar, A. H. Chan
}

\address{Department of Physics, National University of Singapore\\
Block S12 Level 2 Science Drive 2\\
Singapore 117551\\
%%\footnote{State completely without abbreviations, the
%%affiliation and mailing address, including country and e-mail address.
%%Typeset in 8 pt Times Italic.}\\
ang.h.w@u.nus.edu}

\author{M. Rybczy\'nski, Z. W\l odarczyk}

\address{Institute of Physics, Jan Kochanowski University\\
\'Swi\c{e}tokrzyska 15, 25-406 Kielce, Poland
}

\author{G. Wilk}

\address{National Centre for Nuclear Research, Department of Fundamental Research\\
Ho\.za 69, 00-681, Warsaw, Poland}
\maketitle

\pub{Received (Day Month Year)}{Revised (Day Month Year)}

\begin{abstract}
As shown recently, one can obtain additional information from the measured charged particle multiplicity distributions, $P(N)$, by extracting information from the modified combinants, $C_j$. This information is encoded in their specific oscillatory behavior, which can be described only by some combinations of compound distributions such as the Binomial Distribution. This idea has been applied to $pp$ and $p\bar{p}$ processes thus far. In this note we show that an even stronger effect is observed in the $C_j$ deduced from $e^+e^-$ collisions. We present its possible explanation in terms of the Generalised Multiplicity Distribution (GMD) proposed some time ago.

\end{abstract}

\ccode{PACS Nos.: 12.40.Ee}

\section{Introduction}	
Recently it was shown that the measured multiplicity distributions, $P(N)$, contain some additional information on the multiparticle production process, so far undisclosed \cite{JPG, IJMPS, MWW18, MWW1}. The basic idea was to apply the recurrence relation  used in counting statistics when dealing with multiplication effects in point processes \cite{ST}. Its important feature is that it connects all multiplicities by means of some coefficients $C_j$ ({\it modified combinants}), which define the corresponding $P(N)$ in the following way:

\begin{equation}
(N + 1)P(N + 1) = \langle N\rangle \sum^{N}_{j=0} C_j P(N - j). \label{rr2}
\end{equation}
These coefficients contain the memory of the particle $N+1$ about all the $N-j$ previously produced particles and, most important, they can be directly calculated from the experimentally measured $P(N)$ by reversing Eq. (\ref{rr2}) and putting it in the form of the recurrence formula for $C_j$: \cite{JPG}
\begin{equation}
\langle N\rangle C_j = (j+1)\left[ \frac{P(j+1)}{P(0)} \right] - \langle N\rangle \sum^{j-1}_{i=0}C_i \left[ \frac{P(j-i)}{P(0)} \right]. \label{rCj}
\end{equation}
Oscillations were previously observed in the modified combinants derived from $P(N)$ values measured in $pp$ and $p\bar{p}$ experiments, \cite{JPG, MWW1} whereas the $C_j$ values derived from the popular Negative Binomial Distribution (NBD) are monotonically decreasing with increasing $j$. \footnote{The only condition is that the data sample under consideration is large enough, otherwise the oscillatory behavior is washed out by fluctuations.\label{foot_statistics}} On the other hand, the Binomial Distribution (BD) gives strongly oscillating $C_j$ (with period two, not observed in the above data). To fit these data, one needs the BD to be compounded with another distribution that can effectively control the period and amplitude of the resulting oscillations (in Ref \refcite{MWW1} it was compounded with an NBD).

\begin{figure}[t]
\begin{center}
\includegraphics[scale=0.70]{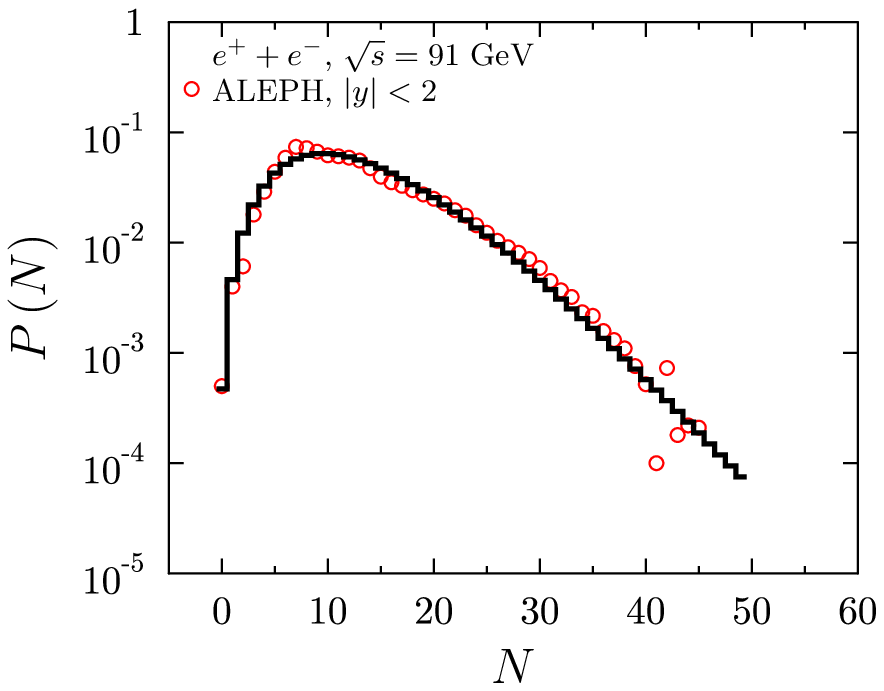}
\includegraphics[scale=0.70]{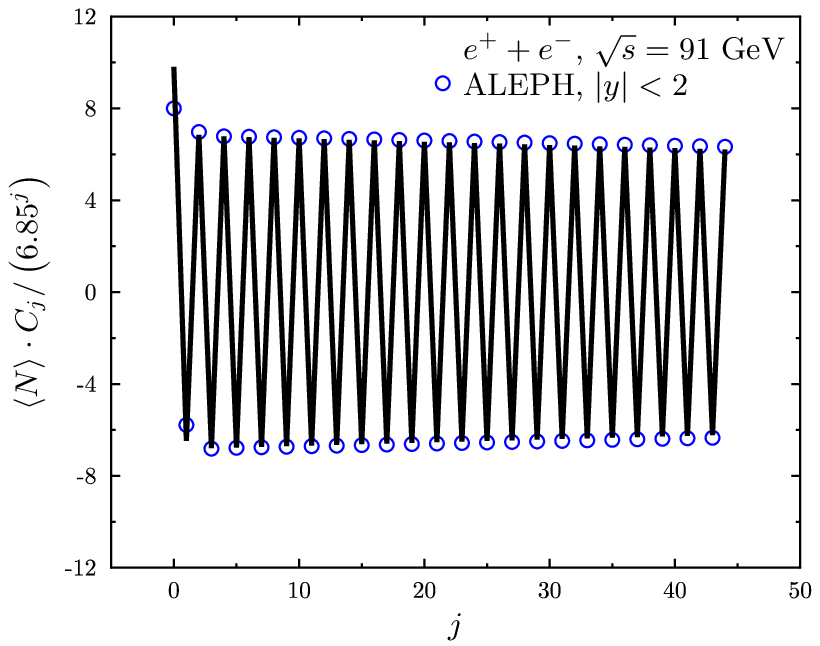}
\end{center}
\vspace{-5mm}
\caption{Left panel: data on $P(N)$ measured in $e^+e^-$ collisions by the ALEPH experiment at $91$ GeV \cite{ALEPH} are fitted by the GMD distribution Eq. (\ref{GMD}) with parameters: $\langle N\rangle = 12.991$, $k = 3.5$ and   $k' = 0.7348$. Right panel: the modified combinants $C_j$ derived from these data (note the significant dependence of the amplitude on rank $j$. The oscillation amplitude of the plot has been scaled accordingly making it possible to plot the results. Otherwise the amplitudes would grow in a power-law fashion). As shown in the following sections, they can be fitted by the $C_j$ obtained from the $P(N)$ derived from GMD.}
\label{GMD-F1}
\end{figure}

It turns out that in the case of multiplicity distributions of charged particles produced in $e^+e^-$ collisions the observed oscillations are much stronger. Fig. \ref{GMD-F1} shows the results for $P(N)$ and for the corresponding $C_j$ deduced from the ALEPH experiment data. \cite{ALEPH} The $C_j$'s can be fitted by using a compound distribution involving the Generalized Multiplicity Distribution (GMD) to be introduced in Section \ref{sec:GMD}.

The modified combinants are closely related to the {\it combinants} $C^{\star}_j$ introduced in Ref. \cite{Combinants} defined in term of the generating function. $G(z)=\sum^{\infty}_{N=0} P(N) z^N $,
\begin{equation}
C^{\star}_j = \frac{1}{j!} \frac{d^j \ln G(z)}{d z^j}\bigg|_{z=0}, \label{C_j_star}
\end{equation}
It can be shown that $C_j$ and $C^{\star}_j$ are related as follows:
\begin{equation}
C_j = \frac{j+1}{\langle N\rangle} C^{\star}_{j+1}, \label{connection}
\end{equation}
By the above relations, $C_j$ can be expressed in terms of the generating function $G(z)$ of $P(N)$ as
\begin{equation}
\langle N\rangle C_j = \frac{1}{j!} \frac{ d^{j+1} \ln G(z)}{d z^{j+1}}\bigg|_{z=0}.
\label{GF_Cj}
\end{equation}
This relation is particularly useful when $C_j$ are calculated from some compound multiplicity distribution defined by a generating function $G(z)$ for which Eq. (\ref{rCj}) would be too difficult to apply \cite{MWW1}.

\section{Generalized Multiplicity Distribution - GMD}\label{sec:GMD}

The GMD was introduced \cite{Chewetal} as an alternative to the NBD solution to study multiplicity distributions. It has the following form:
\begin{equation}
P(N) = \frac{\Gamma(N+k)}{\Gamma(N-k'+1)\Gamma\left( k' + k\right)} p^{N-k'}( 1 - p )^{k+k'},  \label{GMD}
\end{equation}
where
\begin{equation}
p = \frac{\langle N\rangle - k'}{\langle N\rangle + k}. \label{p-GMD}
\end{equation}
The GMD has been successfully applied to $p\bar{p}$ reactions \cite{p-pbar} and $e^+e^-$ annihilation \cite{epem}.
It is based on the stochastic branching equation describing the total multiplicity distribution of partons inside a jet, \cite{GMD}
\begin{eqnarray}
\frac{d P(n)}{d t} &=& -\left( An + \tilde{A}m\right) P(n) + \nonumber\\
&& + A(n - 1) P(n-1) + \tilde{A}m P(n-1), \label{sbe-GMD}
\end{eqnarray}
where
\begin{equation}
t = \frac{6}{11N_c - 2N_f}\ln \left[ \frac{\ln\left( \frac{Q^2}{\mu^2}\right)}{\ln \left( \frac{Q_0^2}{\mu^2}\right)}\right] \label{evolution=parameter}
\end{equation}
is the QCD evolution parameter, with $Q$ denoting the initial parton invariant mass, $Q_0$ the hadronization mass and $\mu$ the QCD mass scale (in GeV). $N_c = 3$ is the number of colours and $N_f=4$ is the number of flavors. $P(n)$ is the probability distribution of $n$ gluons and $m$ quarks in the QCD evolution. In this model, the number of quarks is assumed to be fixed. Parameters $A$ and $\tilde{A}$ denote, respectively, the average probabilities of the $g \to gg$ and $q \to qg$ processes (in the version of Eq. (\ref{sbe-GMD}) used here the contribution of $g \to q\bar{q}$ process has been neglected). The initial number of gluons, $k'$, determines (in the average sense) the initial condition of the generating function, which is $G(t=0,z) = z^{k'}$. The parameter $k = m\tilde{A}/A$ is related (in the average sense) to the initial number of quarks. The Local Parton Hadron Duality \cite{LPHD,LPHD2} was invoked to connect the parton-level results to the experimental data, i.e. the hadron spectra were required to be proportional to the corresponding parton spectra. The entire hadronization process is then parameterized by a single parameter, which determines the overall normalization of the distribution but does not affect moments of order one and above.

The generating function of the GMD is given by
\begin{equation}
G(z) = z^{k'} [ z + (1 - z)\kappa] ^{-(k+k')}, \label{Gz-GMD}
\end{equation}
where
\begin{equation}
\kappa = e^{At}, \label{kappa}
\end{equation}
and the corresponding mean multiplicity is
\begin{equation}
\langle N\rangle = k(\kappa - 1) + \kappa k'. \label{MeanN}
\end{equation}
This can be derived by noting that generating function (\ref{Gz-GMD}) can be also calculated directly using $P(N)$ from Eq. (\ref{GMD}), in which case we obtain
\begin{equation}
G(z) = \sum_{N=k'}^{\infty} z^N P(N) = z^{k'}\left( \frac{1-p}{1-pz}\right)^{k+k'}. \label{CompGz}
\end{equation}
Comparing Eq. (\ref{CompGz}) with Eq. (\ref{Gz-GMD}) one gets that
\begin{equation}
p = 1 - \frac{1}{\kappa} = 1 - e^{-At}. \label{pandkappa}
\end{equation}
Using Eq. (\ref{p-GMD}) for $p$ one gets $\left<N\right>$ in the form of Eq. (\ref{MeanN})~\footnote{For $t=0$ or equivalently $\kappa = 1$ we have $\langle N\rangle = k'$ and $\langle N\rangle$ increases with $Q^2$ as $\langle N\rangle \sim \left( \ln \frac{Q^2}{\mu^2}\right)^{6A/\left(11N_c-2N_f\right)}$.}.

Note that the distribution $P(N)$ described by Eq. (\ref{GMD}) is defined for $N \ge k'$. Hence, both the normalization of the GMD distribution as well as the generating function $G(z)$, Eq. (\ref{Gz-GMD}), are also correspondingly defined for such a range of $N$. This constraint will not affect the values of $C_j$ calculated from Eq. (\ref{rCj}) if $P(0) > 0$. In this instance, only the ratio $P(N)/P(0)$ is required while the normalization cancels out. However, the $C_j$ values calculated from Eq. (\ref{GF_Cj}) using the form of the generating function given by Eq. (\ref{Gz-GMD}) diverge due to the constraint on the range of values $N$ can take.

\section{Normalization of GMD}
\label{sec:N-GMD}

For integer values of $k'$, the GMD distribution (\ref{GMD}) can be interpreted as an NBD "shifted" by $k'$, where $P_{NBD}(N,k)$ denotes the NBD:
\begin{equation}
P_{GMD}\left(N,k,k'\right) = P_{NBD}\left(N-k',k+k'\right), \label{Shifted}
\end{equation}
The normalized form of the GMD as presented in Eq. (\ref{GMD}) has a requirement that $N\geq k'$. However, if we were to extend the domain such that $N\in[0,\infty)$, then the normalized modified GMD becomes

%%The GMD normalization factor $C$ contained in Eq. (\ref{GMD}) is given by:
%%\begin{equation}
%%C = \frac{1}{\Gamma \left(k' + k\right)}p^{-k'}(1 - p)^{k+k'}, \label{Norm}
%%\end{equation}
%%and is valid for $N \ge k'$. It ensures that $\sum_{N=k'}^{\infty} P_{GMD}(N) = 1$.

\begin{equation}
P'_{GMD}(N) = \frac{\Gamma(N+k)}{\Gamma(N-k'+1)}\cdot \frac{\Gamma\left( 1-k'\right)}{\Gamma(k)}\cdot \frac{p^N}{_2F_1\left(1,k,1-k';p\right)},
\label{NormFull}
\end{equation}
where $_2F_1(a,b,c;z)$ is a hypergeometric function. The generating function is then given by
\begin{equation}
G(z) = \sum_{N=0}^{\infty} z^N P'_{GMD}(N) = \frac{_2F_1\left(1, k, 1-k';pz \right)}{_2F_1\left(1,k,1-k';p\right)},
\label{GFF}
\end{equation}

It turns out that when we calculate the modified combinants $C_j$ using the method of Eqn (\ref{GF_Cj}) with Eqn (\ref{GFF}), the resulting $C_j$'s do not diverge. Differentiating the logarithm of Eqn (\ref{GFF}) does not result in a factor of $z$ in the denominator, but rather a polynomial in $z$. The resulting $C_j$ does not diverge due to the presence of the constant term in the polynomial. Nonetheless, for $k' > 1$ the problem still remains\footnote{Note that for small values of $z$ the generating function (\ref{GFF}) is equal to $G(z) \simeq 1 + \frac{kp}{1 - k'} z$, i.e., it is of the BD type and this results in the oscillations characteristic for the BD.}.

We have seen the limitations in a model involving the GMD used in isolation for the purpose of modified combinant analysis. They manifest either in the form of a restricted domain of $N$ (cf. Eq. (\ref{GMD}), or in $k'$ (cf. Eq. (\ref{GFF})). In the next section, we will propose a modification to the GMD to ensure its continued employment in modified combinant analysis, while at the same time circumventing the constraints presented above.

\section{Imprints of acceptance}
\label{sec:imprints}

We shall now propose a modification of the initial $P(N)$ that will allow for $N < k'$. To this end, let us assume that the $P_{GMD}(N)$ as given by Eq. (\ref{GMD}) presents a real distribution which describes the multiplicity in full phase space. However, experimentally, only multiplicity in a limited rapidity window $\Delta y$ is measured. Let us assume therefore that the detection process is a Bernoulli process described by the Binomial Distribution (BD) with the generating function
\begin{equation}
F(z) = 1 - \alpha + \alpha z, \label{G-BD}
\end{equation}
where $\alpha$ denotes the probability of the detection of a particle in the rapidity window $\Delta y$. The number of measured particles $N$ is given by:
\begin{equation}
N = \sum^M_{i=1} n_i, \label{N}
\end{equation}
where $n_i$ follows the BD with the generating function $F(z)$ given by Eq. (\ref{G-BD}) and $M$ comes from the GMD with the generating function $G(z)$ given by Eq. (\ref{Gz-GMD}). The {\it measured multiplicity distribution}, $P(N)$, is therefore given by the GMD compounded with the BD, and its generating function\footnote{ Note that such procedure applied to NBD gives again the NBD with the same $k$ but with modified $p$, which is now equal to $p' = \frac{p\alpha}{ 1 - p + p\alpha}$.} is:

\begin{eqnarray}
\hspace{-3mm}H(z) &=& G[F(z)]\nonumber \\
                  &=& (1-\alpha+\alpha z)^{k'}[1+\alpha (\kappa - 1)-\alpha(\kappa - 1)z]^{-(k+k')}
\label{BD-GMD}
\end{eqnarray}

\begin{figure}[t]
\begin{center}
\includegraphics[scale=0.70]{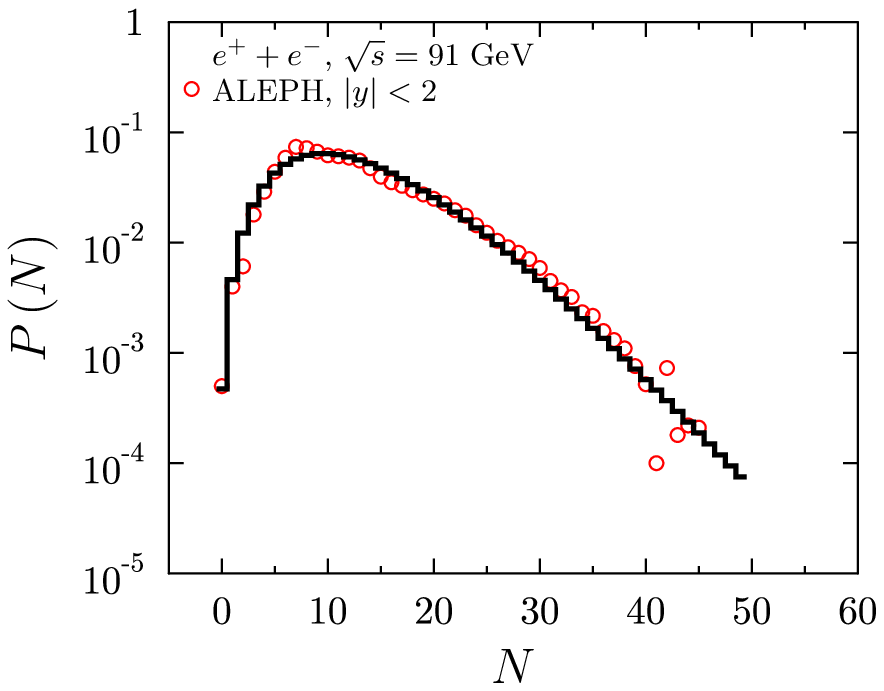}
\includegraphics[scale=0.72]{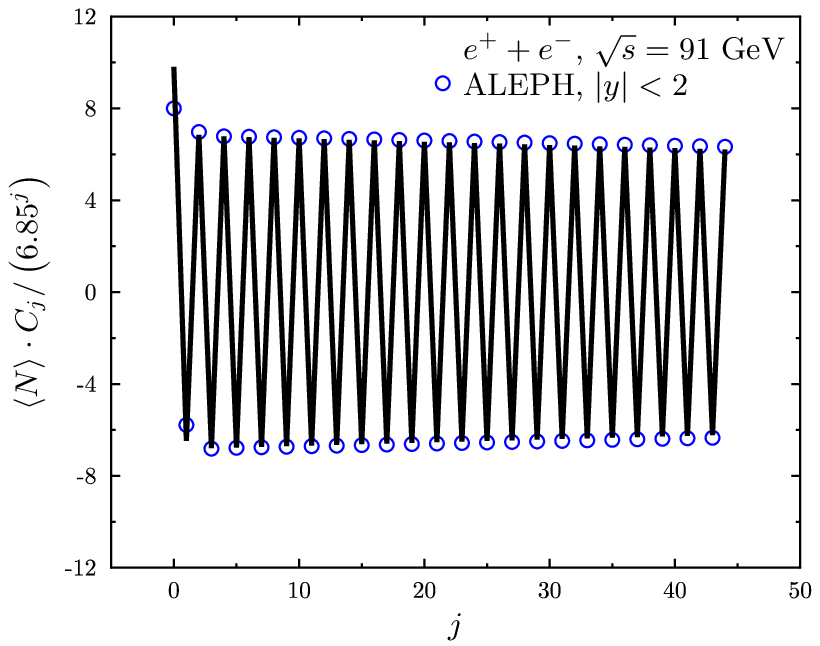}
\end{center}
\vspace{-5mm}
\caption{Left panel: Data on $P(N)$ measured in $e^+e^-$ collisions by the ALEPH experiment at $91$ GeV \cite{ALEPH} are fitted by the distribution obtained from the generating function given by Eq. (\ref{BD-GMD}) with parameters: $\alpha = 0.8725$, $k' = 1$, $k = 3.2$ and $\kappa = 4.585$. Right panel: the modified combinants $C_j$ deduced from these data on $P(N)$ are displayed. They can be fitted by $C_j$ obtained from the same generating function given by Eq. (\ref{BD-GMD}) with the same parameters as used for fitting $P(N)$. }
\label{GMD-F2}
\end{figure}

Note that the generating function (\ref{BD-GMD}) is the product of the generating function of the BD,
\begin{equation}
f(z) = ( 1 - \alpha + \alpha z)^{k'}, \label{G-BD-k'}
\end{equation}
and the generating function of the NBD
\begin{equation}
g(z) = [ 1 + \alpha( \kappa - 1) - \alpha (\kappa - 1 )z]^{-\left( k + k'\right)}. \label{G-NBD}
\end{equation}
Using general Leibniz rule we have that
\begin{eqnarray}
P(N) &=& \frac{1}{N!}\frac{d^N H(z)}{d z^N}\Bigg|_{z=0} \nonumber\\
\!\!\!\!\!&=& \frac{1}{N!}\sum_{i=max\left\{0,N-k'\right\}}^{N} \binom{N}{i}\frac{d^{N-i}f(z)}{dz^{N-i}}\frac{d^ig(z)}{dz^i}\Bigg|_{z=0}. \label{BD-NBD}
\end{eqnarray}
Modified combinants $\langle N\rangle C_j$ calculated  using the generating function (\ref{BD-GMD}) are given by the sum of the respective modified combinants for the BD and the NBD:
\begin{equation}
\langle N\rangle C_j = \frac{1}{j!}\frac{d^{j+1}\ln f(z)}{d z^{j+1}}\Bigg|_{z=0} + \frac{1}{j!}\frac{d^{j+1}\ln g(z)}{d z^{j+1}}\Bigg|_{z=0}. \label{CjBDNBD}
\end{equation}
We can expect therefore oscillations with period equal to $2$, which are superimposed on the monotonically decreasing values:

\begin{eqnarray}
\langle N\rangle C_j &=& (-1)^j k'\left( \frac{\alpha}{1-\alpha} \right)^{j+1} + \nonumber\\
&& +\left( k + k'\right)\left[ \frac{\alpha (\kappa -1)}{1 + \alpha (\kappa - 1)} \right]^{j+1}. \label{CjFin}
\end{eqnarray}
Fig. \ref{GMD-F2} shows this such approach works very well (however, looking on the experimental $C_j$, we can suspect that $C_j$ are increasing for small $j$, this effect has its source in the second term of Eq. (\ref{CjBDNBD})).

\section{Scenario of two sources}\label{sec:2-sources}

\begin{figure}[t]
\begin{center}
\includegraphics[scale=0.72]{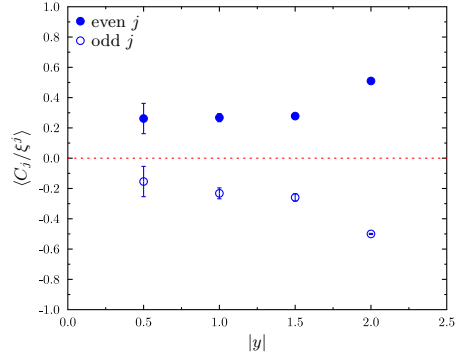}
\end{center}
\vspace{-5mm}
\caption{Mean value $\langle C_j/\xi^j\rangle$  (averaging is performed over available ranks $j$) for even and odd ranks $j$ evaluated from $P(N)$ in different rapidity windows for $\xi = 6.85$, $2.35$, $1.2$ and $0.91$ for $|y|< 2.0$, $1.5$, $1.0$ and $0.5$, respectively.}
\label{GMD-F3}
\end{figure}

Actually, there is yet another way of treating the $e^+e^-$ data. The generating function (\ref{BD-GMD}) can be formally treated as a generating function of the multiplicity distribution $P(N)$ in which $N$ consists of the sum of particles produced from the BD ($N_{BD}$) and the NBD ($N_{NBD}$):

\begin{equation}
N = N_{BD} + N_{NBD}. \label{NN}
\end{equation}
In this case Eq. (\ref{BD-NBD}) can be written as
\begin{equation}
P(N) = \sum_{i=0}^{min\left\{ N,k'\right\}} P_{BD}(i)P_{NBD}(N-i) \label{PbdPnbd}
\end{equation}
and its associated modified combinants Eq. (\ref{CjFin}) can be written as
\begin{equation}
\langle N\rangle C_j = \left< N_{BD}\right>C_j^{(BD)} + \left< N_{NBD}\right> C_j^{(NBD)}. \label{Cjbdnbd}
\end{equation}
The fits shown in Fig. \ref{GMD-F2} correspond to parameters: $k'=1$ and $p'=0.8725$ for the BD and $k=4.2$ and $p=0.75$ for the NBD. It is clear that the implementation of the process in Eq. (\ref{NN}) supports our previous interpretation of GMD being a shifted NBD.

\section{Discussion of Results}
\label{sec:Discussion}

\begin{figure}[t]
\begin{center}
\includegraphics[scale=0.65]{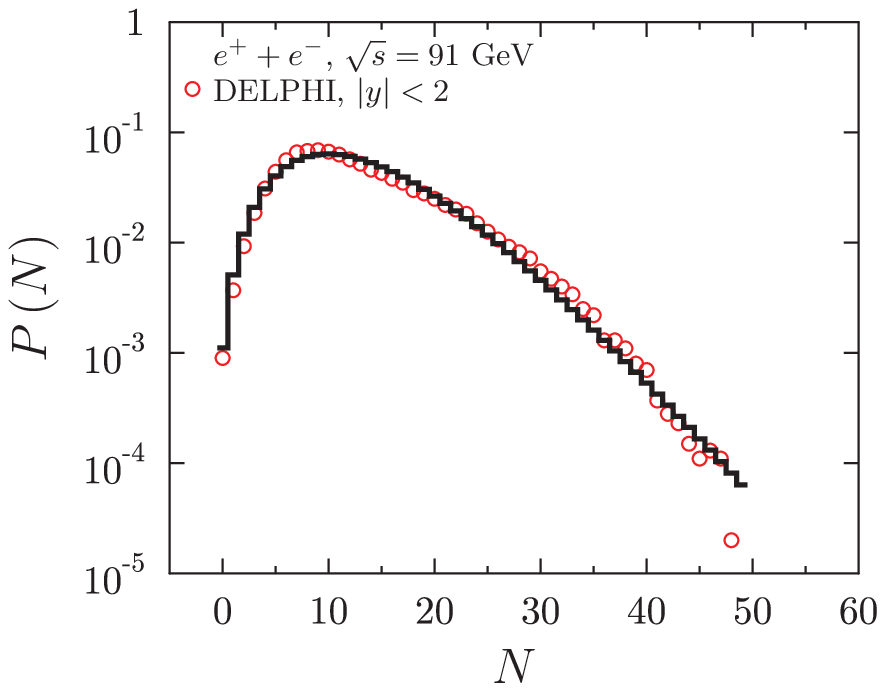}
\includegraphics[scale=0.65]{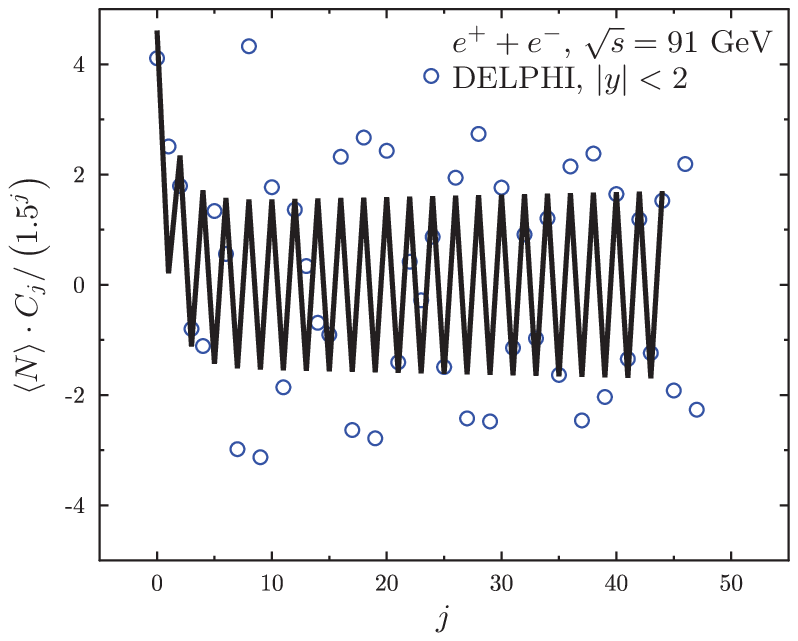}
\end{center}
\vspace{-5mm}
\caption{Left panel: Data on $P(N)$ measured in $e^+e^-$ collisions by DELPHI experiment at $91$ GeV \cite{DELPHI} are fitted by the GMD distribution (\ref{GMD}) with parameters: $\langle N\rangle = 13.1043$, $k = 4.5$ and   $k' = 0.29$. Right panel: the modified combinants $C_j$ deduced from these data on $P(N)$ are displayed. Note that in this case they are too spread to be of any further use. This is clear when comparing them with the respective $C_j$ obtained from the same GMD model with the same parameters as used for fitting of $P(N)$. The reason is too low statistics in DELPHI experiment (cf. text for details).}
\label{GMD-F4}
\end{figure}

Section \ref{sec:N-GMD} highlights the difficulties in deriving the modified combinants of the GMD when used by itself. The proposed alternatives to circumvent them are then presented in Section \ref{sec:imprints}. The $C_j$ resulting from the original GMD diverges due to a factor of $z^{k'}$ in the denominator. This factor has its origins in the assumed initial condition of $k' = const$. Had the initial condition taken the form of a binomial distribution \cite{Wilk.Stochastic.Background} the resulting generating function will match that in Eqn (\ref{BD-GMD}) with none of the restrictions encountered in Sect \ref{sec:N-GMD}.

Oscillatory behaviour of modified combinants is observed for all $P(N)$ measured by ALEPH in different rapidity windows. In Fig. \ref{GMD-F3} we showed the mean values of the amplitudes of oscillations (rescaled by factor $\xi^j$) for even and odd values of $j$,  evaluated from the experimental multiplicity distributions $P(N)$ measured in various rapidity windows $|y|<2.0$, $1.5$, $1.0$ and  $0.5$.  The averages for both even and odd $j$'s are distinctly different from zero, supporting the fact the oscillations do not have their origins in random fluctuations (in which case both values should independently oscillate around zero).

It must be noted that had the DELPHI data \cite{DELPHI} been used, the analysis of the corresponding $C_j$'s would not have been possible. This is illustrated in Fig. \ref{GMD-F4} where the experimentally obtained modified combinants $C_j$ are too scattered to discern any underlying oscillations due to the low statistics of the DELPHI data (cf footnote \ref{foot_statistics}). The measured $P(0)$ from DELPHI is also about twice as large as that from ALEPH. This explains the much smaller oscillation amplitude seen in Fig \ref{GMD-F4} (cf. Eqn (\ref{rCj})). For comparison, the data from ALEPH was obtained over a total of $3\cdot 10^5$ data events, 5 times that of the DELPHI data set. The difference in the unfolding procedure applied to measured data between ALEPH and DELPHI contributes partly to the more scattered $C_j$ data in Fig \ref{GMD-F4}. Nevertheless, the derived $C_j$ from both ALEPH and DELPHI exhibits oscillations of period 2, which appears characteristic of the experimental data used in this paper.

Different period of oscillations in $C_j$ results from different reaction types. In the $e^+e^-$ annihilation data presented, the derived $C_j$ oscillates with period of 2. On the other hand, $pp$ collisions give $C_j$ oscillating with a longer period on the order of 10. The period of $C_j$ oscillation has been previously studied \cite{JPG, MWW1} under the framework of compound distributions. The oscillatory behaviour is controlled by the BD component while the NBD component controls the amplitude and period. When interpreted in the framework of cluster production, the number of clusters follow the BD while the number of particles per cluster obeys the NBD. The period of $C_j$ oscillation has been shown to be related to the mean multiplicity of the NBD.

\section{Concluding remarks}
\label{sec:Sum}

In our paper, we have demonstrated that modified combinants are valuable additional tools for the examination between different models, in conjunction with the more traditional approach of analysis of multiplicity distribution. A careful analysis of the modified combinants $C_j$ deduced from the experimentally measured $P(N)$ can provide additional information on the dynamics of particle production, thereby allowing us to reduce the number of potential explanations presented amongst the various models. In this context, $C_j$ deduced from experimental measurements of $P(N)$ becomes an important addition to the toolkit in the field of particle physics phenomenology.

A detailed discussion of the sensitivity of modified combinants $C_j$ to statistics of events and the associated uncertainties of measurements is given in Ref. \refcite{MWW1} (see also Ref. (\cite{Zborovsky})). Notwithstanding the sensitivity of oscillations of modified combinants to systematic uncertainties in the measurements of $P(N)$, the oscillatory signal observed in the modified conbinants derived from ALEPH data is shown to be statistically significant. The oscillations are not artefacts of random fluctuations as seen in Fig. \ref{GMD-F3}. These observations justify the study of oscillations in $C_j$ to discern finer details in experimentally measured multiplicity distributions.

There has been a large number of papers suggesting universality in the mechanisms of hadron production in $e^+e^-$ annihilation, $pp$ and $p\bar{p}$ collisions. Such universality arises from observations of the average multiplicities and relative dispersions in the different types of processes \cite{Biswas,CU,Kittel,G-OR, Bzdak} (cf. for example, Sect. $19$ of Ref. \cite{PDGbook} and references therein). A detailed analysis of the modified combinants derived from the experimental $P(n)$'s reveals differences between the various processes. In $e^+e^-$ annihilation, the $C_j$'s oscillate with a period of 2 with amplitudes increasing as a power-law. On the other hand, $pp$ and $p\bar{p}$ collisions produce $C_j$'s oscillating with approximately 10 times the period of their $e^+e^-$ counterparts, with decaying amplitudes\cite{JPG, MWW1}. Further analysis in this aspect will be welcome.

In what concerns the $e^+e^-$ results discussed here, the most plausible interpretation lies with the GMD approach (with some modifications discussed above). The GMD contains in its structure a BD, both in the original version resulting in Fig. \ref{GMD-F1} and in modified version resulting in Fig. \ref{GMD-F2}. In such cases, any distributions will show oscillatory behaviour in the derived $C_j$'s when compounded with the BD\cite{MWW1}. However, this problem seems at the moment still open and subject to future investigations.

\section*{Acknowledgments}

This research  was supported in part by the National Science Center (NCN) under contracts 2016/23/B/ST2/00692 (MR) and 2016/22/M/ST2/00176 (GW). M.Ghaffar would like to thank NUS for the hospitality where part of this work was done. H.W. Ang would like to thank the NUS Research Scholarship for supporting this study. We would also like to thank Q. Leong for his critical discussion and Dr Enrico Sessolo for reading the manuscript.

%\bibliographystyle{ws-mpla}
%\bibliography{sample}

\end{document}